\documentclass[aip,amsmath,onecolumn,11pt,jcp]{revtex4-1}
\usepackage{bm}%
\usepackage[colorlinks=true,linkcolor=blue]{hyperref}%
\usepackage{graphicx}
\usepackage{dcolumn}
\usepackage{subfigure}
\usepackage{amsmath}
\usepackage{amsfonts}
\usepackage{ragged2e}
\usepackage{amssymb}
\usepackage{geometry}
\usepackage{graphicx}
\usepackage{xcolor}

\usepackage[font=footnotesize,labelfont=bf, justification=raggedright]{caption}

\expandafter\ifx\csname package@font\endcsname\relax\else
 \expandafter\expandafter
 \expandafter\usepackage
 \expandafter\expandafter
 \expandafter{\csname package@font\endcsname}%
\fi
\begin{document}
	
	\title{On Machine Learning Force Fields for Metallic Nanoparticles}

	\author{Claudio Zeni`} 
	\email{claudio.zeni@kcl.ac.uk}
	\affiliation{Department of Physics, King's College London, Strand, London WC2R 2LS, United Kingdom}
	
	\author{Kevin Rossi} 
	\email{kevin.rossi@epfl.ch}
	\affiliation{Laboratory of Computational Science and Modelling, \'Ecole polytechnique f\'ed\'erale de Lausanne, Route Cantonale, 1015, Switzerland}
	\affiliation{Department of Physics, King's College London, Strand, London WC2R 2LS, United Kingdom}
	
	\author{Aldo Glielmo}
	\affiliation{Department of Physics, King's College London, Strand, London WC2R 2LS, United Kingdom}

	\author{Francesca Baletto}
	\email{francesca.baletto@kcl.ac.uk}
	\affiliation{Department of Physics, King's College London, Strand, London WC2R 2LS, United Kingdom}
\begin{abstract}
Machine learning algorithms have recently emerged as a tool to generate force fields which display accuracies approaching the ones of the \emph{ab-initio} calculations they are trained on, but are much faster to compute.
The enhanced computational speed of machine learning force fields results key for modelling metallic nanoparticles, as their fluxionality and multi-funneled energy landscape needs to be sampled over long time scales. 
In this review, we first formally introduce the most commonly used machine learning algorithms for force field generation, briefly outlining their structure and properties.
We then address the core issue of training database selection, reporting methodologies both already used and yet unused in literature.
We finally report and discuss the recent literature regarding machine learning force fields to sample the energy landscape and study the catalytic activity of metallic nanoparticles.
\begin{figure*}[h!]
\begin{center}
\includegraphics[width=0.53\columnwidth]{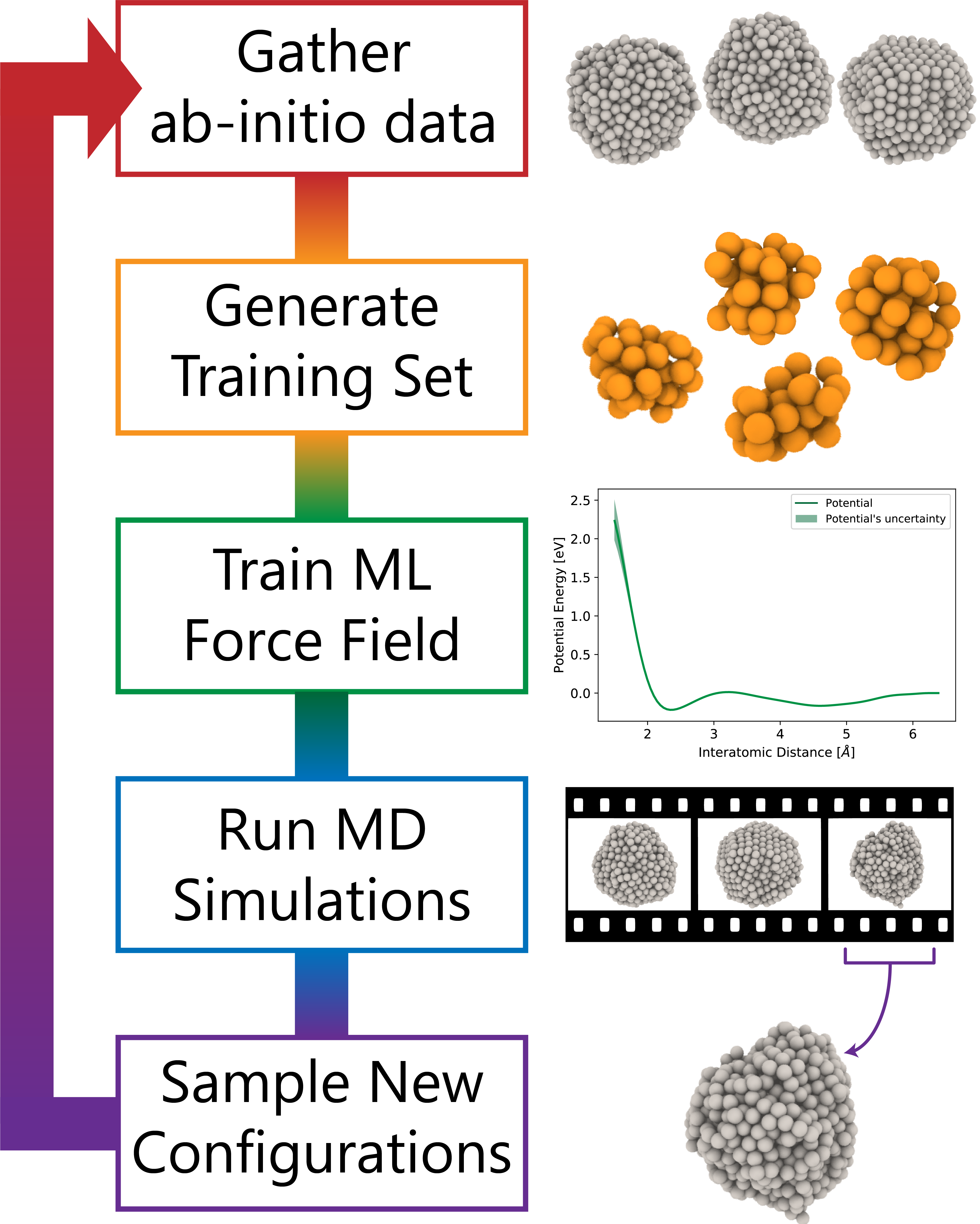}
\end{center}
\end{figure*}

\end{abstract}
\maketitle

\section{Introduction}

Metallic nanoparticles (MNPs) are finite objects, containing up to several thousands of atoms.
Being 0D, MNPs can display a very diverse and complex configurational space with many local minima, often very close in energy. \cite{Baletto2005, Wang2012, Foster2018}  
MNPs' chemical and physical properties - different from their single atom or bulk counterparts - depend strongly on the MNPs' architecture (i.e., their size, shape, chemical composition and chemical ordering).
Therefore, harnessing the full potential of MNPs for practical applications in catalysis, plasmonics, optics, memory-storage, and biomedicine, hinges on the understanding and control of the dependence between the variety of accessible/available isomers over time and hence of MNPs' properties. \cite{Sun2007,Rossi2019,Noguez2007,Cuenya2010,Giljohann2010}
 \\
In the past two decades many efforts have been carried out to sample the multi-funnelled energy landscape of MNPs, mainly to identify MNPs' putative global minima. \cite{Ferrando2008}
To this end, various sampling methods have been employed, such as Monte Carlo schemes \cite{Rahm2017}, basin hopping \cite{Li1987, Chen1998, Rossi2009, Barcaro2006, Calvo2016}, hyperspatial optimization \cite{Pickard2019}, and evolutionary algorithms \cite{Johnston2003, Heiles2013, Lazauskas2017, Jian2014, Lepeshkin2018} to name a few.
\\
 While prediction from global minimum or archetypical structures may be helpful in deciphering structure-property relationship, only the accurate estimate of the finite-temperature population of relevant isomer enables to obtain accurate predictions of the ensemble properties.
 From a numerical point of view, molecular dynamics, path sampling \cite{Trygubenko2006, Koslover2007}, metadynamics \cite{Laio2002, Laio2008, Pavan2015, Gould2016, Rossi2017}, and finite temperature partition function evaluation approaches \cite{Skilling2006, Partay2010, Rossi2018, Dorrell2019, Doye1996} offer a dramatically effective way to account for structural changes and fluxional behaviours, and thus to estimate faithful population distributions.
 The evaluation of converged results is however expensive, i.e., it necessitates of thousands if not millions of energy and force calculations.\cite{Gould2016,  Rossi2018}
\\
In the recent years the advent of machine learning (ML) techniques paved the way towards establishing a novel paradigm in the computational studies of nanoparticles' structural and chemophysical properties.
Together with big-data techniques that easily and better characterize nanoparticles' geometries during simulations \cite{Fernandez2015,Yan2018,Gasparotto2018} and   experiments,\cite{Timoshenko2017,Timoshenko2019,Madsen2018} machine learning force fields (ML-FFs) represent the most ground-breaking application of artificial intelligence models to further the breath and scope of numerical prediction and design of MNPs.
\\
Machine learning force fields' (FFs) accuracy is comparable to the one of the electronic structure methods they are trained on, but their speed can approach the one of semi-empirical methods.
Their use in algorithms to sample minima and transition states in nanoparticles' energy landscapes thus promises to reduce the computational cost of sampling schemes and in turn to establish them as routine, comprehensive and exhaustive procedures that enable to bridge simulations and real systems used in experimental investigations.
\\
The aim of this review is to offer a representative and pedagogical overview of the state-of-the-art of ML-FFs for MNPs, while also presenting a discussion on the open questions related to the subject.
The review is organized as follows:
in section \ref{sec:descriptors} the local atomic environment representations most commonly employed in ML-FF generation will be discussed and presented to the reader.
In section \ref{sec:algorithms} three supervised ML algorithms for ML-FF generation will be formally introduced and briefly explained: linear regression, artificial neural networks (ANN), and Gaussian process regression (GPR).
Section \ref{sec:database} will discuss the issue of training database selection, presenting some methods that have been already used in literature together with algorithms that have not already been used in the field.
Subsequently, section \ref{sec:examples} will present examples from literature on the application of ML to the study of nanoparticles, together with figures extracted from selected works.
Finally, the conclusion will summarise the earlier presented discussion. 

\section{Descriptors of local atomic environments}
\label{sec:descriptors}
A force field establishes a mapping between atomic coordinates and the corresponding total energy of the system.
The construction of a force field can therefore be framed as a supervised learning problem where the objective is to find a function that yields a total energy $E$ (and/or forces $\mathbf{F}_i$) given the coordinates and species of atoms in a system $\mathcal{R}$.
The total energy $E(\mathcal{R})$ of the system $\mathcal{R}$ can be learnt as a function of the global coordinates of the system, but this leads to computational costs which scale non-linearly with the size of the system. \cite{Sumpter1992, Blank1995, Gassner1998}
Therefore, in order to improve the computational scaling, and to construct force fields which can be used on systems with arbitrary number of atoms, the total energy $E(\mathcal{R})$ of a system can be approximated as a sum of local energy contributions $\epsilon(\rho_i)$, each pertaining to an atom $i$ surrounded by its local atomic environment $\rho_i$: \cite{Behler:2007fe}
\begin{equation}
E(\mathcal{R}) \approx \sum_{i \in \mathcal{R}} \epsilon(\rho_i),
\label{eq::global_energy}
\end{equation}
where the local atomic environment $\rho_i$ refers to a vector or higher dimensional object that contains information regarding the positions and atomic species of atoms contained in a sphere of radius $r_c$ centred on atom $i$.
The approximation of Eq. \ref{eq::global_energy} is based on the \emph{near-sightedness} principle, i.e. on the fact that forces acting on atoms are not strongly dependant on the positions and species of atoms very far away. \cite{Kohn1996}
The \emph{near-sightedness} assumption does not hold for interactions which are inherently long-ranged, i.e. electrostatic interactions; and these have to be treated separately.
\\
The force vectors $\mathbf{F}_i$ can be calculated using the definition:
\begin{equation}
\mathbf{F}_{i} = - \dfrac{\partial E}{\partial \mathbf{r}_{i}}
\label{eq::force_derivative}
\end{equation}
where $\mathbf{r}_{i}$ is the vector containing the Cartesian coordinates of atom $i$.
\\
A data set $\mathcal{D}$ composed of $N$ input-output pairs $\mathcal{D} = \{ \mathcal{R}_n, E_n \} $, with $n \in \{1, \cdots, N \}$ is required for supervised learning problems.
A fraction of such data set is used to train the model (training set $\mathcal{D}_{tr}$), and the remaining part to test its accuracy (validation set $\mathcal{D}_{val}$).
\\
We will now assume the learning problem to be focused on predicting local energies $\epsilon(\rho_i)$, since total energies and forces can be derived from local energies (Eqs. \ref{eq::global_energy},  \ref{eq::force_derivative}).
This assumption is made mainly for educational purposes, as machine learning force fields can be trained using total energies, forces, or a combination of the two. 
\\
In general, a local energy $\epsilon(\rho_i)$ of a given atom $i$ is not learned directly as a function of the atomic Cartesian coordinates. 
On the contrary, the coordinates of the atoms in the vicinity of atom $i$ (within a chosen cutoff $r_c$) are usually first transformed to a vector $\mathbf{q}_i$, commonly called a \emph{descriptor}, and then a regression model is used to learn the map $\epsilon(\mathbf{q}_i)$ from the descriptor to the sought local energy function.
This procedure is highly beneficial to the transferability of the generated force field as well as to the learning speed of the ML algorithm used, provided that the chosen descriptor possesses some key properties. 

\subsection{Descriptor's properties}
A suitable representation of the local environment around an atom is encoded in a descriptor, which must possess the following key properties.
Firstly, in order to strictly constraint the learned force field to respect fundamental physical invariances, any descriptor should be invariant upon rigid translations, rotations and reflections of the local environment, as well as invariant upon any permutation of atoms of the same chemical species.
Arguably, invariance properties could be learned automatically by any sufficiently flexible algorithm when provided with enough data.
Even so, their strict imposition through an invariant representation has been found to be extremely beneficial both for the learning speed and for the transferability of the force field.
\cite{Botu:2014kc,Botu:2015kb,Glielmo2017,Li:2015eb}
Secondly, it is fundamental for a descriptor to be differentiable with respect to the atomic coordinates, as this is required for an analytic computation of the atomic forces after the interpolation of an energy function. 
Thirdly, a descriptor should contain enough information to well capture the relevant physics of the system.
As an illustrative example, the total mass of a local atomic environment $\rho_i$ is a perfectly invariant function, but can hardly provide enough information for constructing a useful interaction model. 
Moreover, it would be desirable for a descriptor not only to contain enough information, but also to capture, to some extent, important features of the sought energy or force function as this will make learning process faster and resulting force field potentially more accurate.
Finally, in addition to all other properties a descriptor should also be computationally inexpensive w.r.t. the reference method (e.g. DFT).
Indeed, as an extreme example, the density functional theory (DFT) force predicted on a given atom fulfils all the conditions given above while still being a useless descriptor in practice.
\\
The requirements listed above leave a large freedom of choosing a single descriptor among countless options.
For this reason, many local atomic environment descriptors have been developed and applied to build ML-FFs \cite{Shapeev2016, Tian2018, Isayev2017}, and in some occurrences such descriptors have been left for a ML algorithm to learn and optimise \cite{Schutt2017}
In the following two paragraphs, we focus on the two descriptors that have been used the most for building atomistic force fields for nanoclusters: atomic symmetry functions \cite{Behler:2007fe} and power spectrum  \cite{Bartok2013}.
Afterwards, we will point out the common features possessed by these successful descriptors. 
Their remarkable similarity will then naturally bring to the exploration of another class of descriptors, simply encompassing explicit $n$-body degrees of freedom, which have recently been used successfully to build force fields for nanoclusters \cite{Zeni2018}.
\\

\subsection{Atomic symmetry functions}
The first descriptor used in the context of ML-FF fitting consists in a set of functions of the local environment which are invariant by construction over the physical symmetries mentioned earlier.
These functions are called atomic symmetry functions (ASF) and, since they were introduced by Behler and Parrinello in their pioneer work on neural networks force fields \cite{Behler:2007fe}, they are also often referred to as Behler-Parrinello functions.
ASFs form a basis set for the expansion of 2-body (radial) and 3-body (angular) distribution functions. \cite{Behler:2007fe}
Over the years, variations on the structure of ASFs have been developed, e.g. using Chebyshev polynomials to improve their spatial resolution's efficiency \cite{Artrith2017}.
A brief description of their initial formulation \cite{Behler:2007fe} follows.
\\
Given a local atomic environment $\rho_i$ defined by the positions $\{ \mathbf{r}_{ij} \}_{j=1}^M$ of $M$ atoms within a cutoff radius $r_c$ of a central atom located at the origin of the reference frame, the 2 and 3-body ASFs take the form, respectively:
\begin{align} 
	G^i_2 &= \sum_{j \in \rho} e^{-\eta (r_{ij} - r_s)^2}f_c(r_{ij})  
\label{eq:ASF2}
\\
	G^i_3 &= \sum_{j \neq k \in \rho} 2^{1-\zeta} (1 + \lambda \cos \theta_{jk})^{\zeta} \, e^{-\eta (r_{ij}^2 + r_{ik}^2 + r_{jk}^2)} f_c(r_{ij}) f_c(r_{ik}) f_c(r_{jk}).
\label{eq:ASF}
\end{align}
where $f_c(r)$ is a cutoff function smoothly approaching zero as $r_{ij}$ approaches the cutoff radius $r_c$. 
Its presence in the symmetry functions ensures that the descriptor smoothly varies when atoms transit through the radial cutoff, thus avoiding any discontinuity in the learned energy or force function.
\\
Both $G^i_2$ and $G^i_3$ depend on some parameters, namely $\boldsymbol{\theta}_2 = (\eta, r_s)$ and $\boldsymbol{\theta}_3 = (\eta, \zeta, \lambda)$ (with $\lambda \in \{ -1, 1 \}$), and the descriptor vector $\mathbf{q}^i_{ASF}$ is built by evaluating these functions (Eqs. \ref{eq:ASF2},  \ref{eq:ASF}) for a large number of these parameters. 
The ASF descriptor, therefore, requires a selection of relevant parameters for the $\boldsymbol{\theta}_2$ and $\boldsymbol{\theta}_3$; this can be achieved via fingerprint optimisations as recently discussed by \textcite{Imbalzano2018}.
\\
The function $G^i_2$ can be seen to provide information about all the distances from the central atom to its $M$ neighbours.
On the other hand, $G^i_3$ encompasses information on every triplet of atoms which include the central atom, meaning that they capture respectively 2 and 3-body information about the local environment $\rho_i$.
It is important to note that using a 2 or 3-body descriptor does not limit the generated ML-FF to be low order, since an arbitrary non linear function of such a descriptor will be able to model higher order interactions as well. \cite{Glielmo2018}
\\
The relative simplicity of implementation, the low computation cost, and perhaps the large body of research demonstrating the effectiveness of this descriptor, have made it the preferred choice for many practical applications in nanoparticle science. 
\cite{Artrith2012,  Artrith2013,Artrith2014a,Artrith2015,Chiriki2016,Chiriki2017,Chiriki2016, Ulissi2017, Ouyang2015, Chiriki2017b, Kolsbjerg2018, Sun2018, Hajinazar2019}
\\

\subsection{Spherical harmonics power spectrum}
The power spectrum of a spherical harmonics (SH) expansion is another popular choice for atomic descriptors. 
It was first introduced to fit energies with GPR \cite{Bartok2013,Bartok:2015iw}, but has later found applications also for ANN FFs \cite{Jindal2017,Jindal2018}.
The power spectrum descriptor has been proven to be equivalent to the widely used ``Smooth Overlap of Atomic Orbitals" (SOAP) representation when using a dot product kernel within GPR.  \cite{Bartok2013}
\\
To build a power spectrum descriptor, it is customary to first write the local environment as a sum of Gaussian functions, each centred on one of the $M$ atoms within the local environment $\rho_i$ defined by the cutoff radius $r_c$:
\begin{equation}
	\rho_i(\mathbf{r}) = \sum_{j \in \rho_i} e^{-(\mathbf{r} - \mathbf{r}_{ij})^2/2\sigma^2}.
\label{eq:gaussian_descr}
\end{equation}
Then the key idea is that of expanding the angular part the above function in a SH basis $\{ Y_{lm}(\hat{\mathbf{r}}) \}$, in order to easily build a rotationally invariant descriptor as the power spectrum of this expansion. \cite{Bartok2013} 
Since SH can only retain the angular information of the above function (Eq. \ref{eq:gaussian_descr}), the radial part is expanded in another basis, $g_n(r)$.
The specific choice of the radial basis is not crucial to the descriptor, common choices are polynomials or equispaced Gaussian functions.
All together, the function $\rho_i(\mathbf{r})$ can hence be formally rewritten as:
\begin{equation}
	\rho_i(\mathbf{r}) = \sum_{n=0}^{\infty}\sum_{l=0}^{\infty}\sum_{m=-l}^{l} c^i_{nlm} \, g_n(r) Y_{lm}(\hat{\mathbf{r}}).
\label{eq:sh_descr}
\end{equation}
The coefficients $c^i_{nlm}$ of the above SH expansion are not rotationally invariant.
However, the power spectrum of the expansion can easily be written down in terms of the SH coefficients as:
\begin{equation}
	p^{i}_{nn'l} = \frac{1}{\sqrt{2l +1}} \sum_{m= -l}^{l} c^i_{nlm} c^{i*}_{n'lm}.
\label{eq:power_spectrum}
\end{equation}
The power spectrum descriptor $\mathbf{q}^i_{PS}$ is defined as the vector containing the power spectrum coefficients described above. It is rotationally invariant and can hence be effectively used to represent atomic environments.
The power spectrum is more expensive to compute than the ASF descriptor, but it requires the user to choose only the width $\sigma$ of the Gaussian functions in Eq.~\ref{eq:gaussian_descr} and where to truncate the expansion in Eq.~\ref{eq:sh_descr}, instead of the full range of ASF parameters.
Recently, a more efficient variant of the power spectrum descriptor which reduces its computational cost has been proposed.
The interested reader is referred to \textcite{Thompson2016} for more detailed information on such variant.
\\

\subsection{Explicit $n$-body features}
In spite of the apparent differences between the mathematical expressions of the two descriptors discussed above, the power spectrum (Eq. \ref{eq:power_spectrum}) and the ASF representation (Eqs. \ref{eq:ASF2} and  \ref{eq:ASF}) are very similar at a fundamental level. 
Indeed, both can be considered as symmetric representations of the 2 or 3-body information present in the atomic environment.
This is particularly clear in the case of the ASFs (Eq. \ref{eq:ASF}), which are seen to depend rather explicitly on distances and angles. 
The $n$-body nature of the power spectrum descriptor is instead less obvious from its mathematical formulation (Eqs. \ref{eq:sh_descr} and \ref{eq:power_spectrum}), but it follows from the analysis reported in Ref.\cite{Glielmo2018}--later formulated also in Ref.\cite{Willatt2019}--that the power spectrum is a 3-body descriptor, meaning that any linear regression built on such a descriptor will give rise to a 3-body force field.
This fact can be intuitively understood considering that the Gaussian expansion in Eq.~\ref{eq:gaussian_descr} is made of contributions coming from pairs of atoms, and can thus be seen as a 2-body descriptors.
The power spectrum is then constructed by mixing rotationally invariant information coming from pairs of such 2-body descriptors (Eq.~\ref{eq:power_spectrum}), making the final descriptor 3-body.
\\
It is not a surprising fact that both of the descriptors mentioned so far (which are indeed among the most used in practical applications) are constructed using 2 and/or 3-body features.
Low order descriptors can in fact be expected to capture well the ionic and covalent nature of chemical bonds, thus representing a rather natural choice. 
Moreover, while the absence of any angular information in a 2-body descriptor will always prevent a correct characterization of higher order interactions, a 3-body descriptor does not present the same problem, meaning that a nonlinear function of a 3-body descriptor (e.g., a neural network or a Gaussian process model) can in principle capture well also higher order interactions. 
\\
Building on these ideas, one can think of explicitly using the 2 and 3-body degrees of freedom present in a given environment \emph{directly} as descriptors. 
For instance, a 2-body descriptor can be simply given by the \emph{unordered} set of distances from the central atom to all other atoms within the configurations, 
\begin{equation}
	\mathbf{q}^i_2 = \{ r_{ij} \}_{j\in \rho_i},
\label{eq:2bdescr}
\end{equation}
while a 3-body descriptor can be given by the unordered set of triplets of distances between a central atom and any two neighbors
\begin{equation}
	\mathbf{q}^i_3 = \{ (r_{ij}, r_{ik}, r_{jk}) \}_{j,k \in \rho_i}.
\label{eq:3bdescr}
\end{equation}
The above descriptors contain by construction the full 2- and 3-body information about the local environment. 
They are moreover computationally efficient, simple to interpret, and do not require any choice of parameters or truncation approximation.
These advantages come at a cost. 
Indeed, the requirement that the above sets are unordered is strictly needed to preserve permutational invariance, and such a condition must be imposed to the ML algorithm processing the inputs.
This can be done rather straightforwardly when using a Gaussian process regression model (see Section \ref{sec:GPR}), and in fact explicit $n$-body features as the one provided above have been extensively used in this framework \cite{Deringer2017, Glielmo2018, Vandermause2019}, also for building force fields for nanoclusters \cite{Zeni2018}.
%

\section{Machine learning algorithms to generate force fields}
\label{sec:algorithms}%
\begin{figure}[h!]
\begin{center}
\includegraphics[width=0.98\columnwidth]{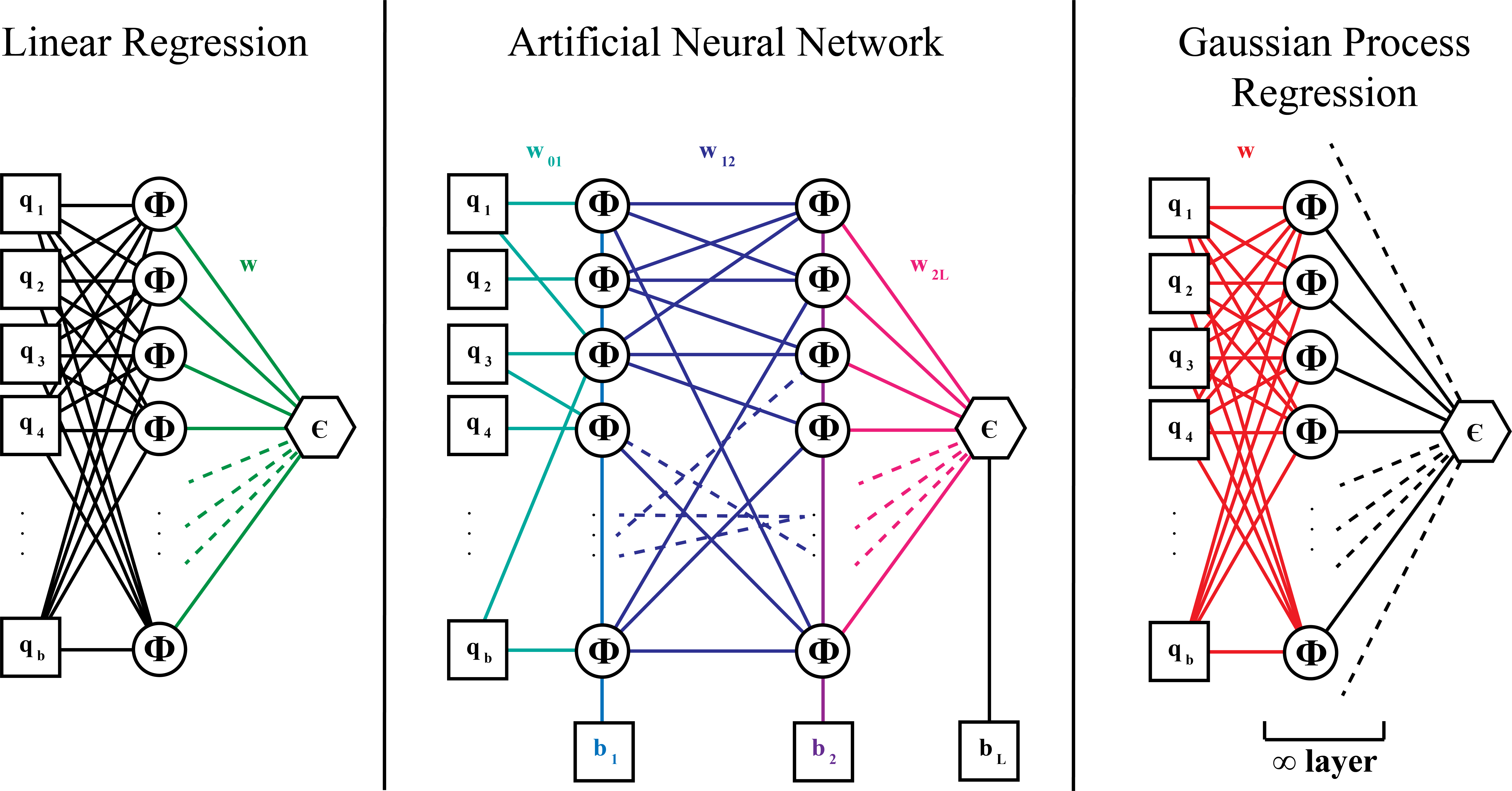}
\caption{Schematic comparison of linear regression, 2-layer  ANN, and GPR for the local energy prediction task.
The symbols in the figure mirror the ones used in the main text.
All three algorithms have been depicted in a similar manner so to ease the recognition of parallellisms and differences between the methods.
Gaussian processes can be imagined to be equivalent to a
fully connected ANN with a single infinite layer, this analogy has been proven rigorously in Refs.  \textcite{neal2012bayesian} and \textcite{Rasmussen:2006vz}.}
\label{fig:regression_comparison}
\end{center}
\end{figure}

The three most commonly used  methods to learn the local energy function $\epsilon(\mathbf{q}_i)$ are linear regression, artificial neural networks and Gaussian process regression.
These methods are schematically compared in Figure \ref{fig:regression_comparison}.
In the following, these three algorithms are briefly presented.
\subsection{Linear Regression}
\label{sec:LR}
A straightforward way to model the local energy function $\epsilon_i$ is via linear regression:
\begin{equation}
\hat{\epsilon}(\mathbf{q}_i) = \mathbf{w}^T \boldsymbol{\phi}(\mathbf{q}_i),
\label{eq::linear_regression}
\end{equation}
where $\mathbf{w}$ are the weights, $\hat{\epsilon}(\mathbf{q}_i)$ indicates the modelled energy function and $\boldsymbol{\phi}$ is a function chosen \emph{a priori} (e.g. polynomial function, sinusoidal function, etc.).
The weights that minimize the squared error loss $L_2$: 
\begin{equation}
L_2 = \sum_{n =1}^{N_{tr}} ||\hat{E}_n - E_n||^2
\label{eq:loss}
\end{equation}
 can be found analytically.
In Eq. \ref{eq:loss} the predicted and real total energies $\hat{E}_n$, $E_n$ of each system are calculated per Eq. \ref{eq::global_energy}, and $N_{tr}$ indicates the number of training points used.
In the linear regression case, the solution to the learning problem is fast to compute, and the predictions the model yields are computationally cheap. 
Despite the simplicity of the learning model, a variation of this method has been proven to yield interesting results, for example in aiding the prediction of adsorption energies for RhAu nanoparticles. \cite{Jinnouchi2017a, Jinnouchi2017b}
\\
Nonetheless, in most applications linear regression models lack the complexity needed to construct an accurate force field.
%
%

\subsection{Artificial Neural Networks}
\label{sec:ANN}
\paragraph{Structure}
ANNs are composed of nodes organised in layers which connect the input $\mathbf{q}_i$ to an output $\epsilon(\mathbf{q}_i)$.
Networks with few hidden layers are called ``shallow'', whereas the term ``deep'' is used to indicate networks with a high number of hidden layers.
Nodes are connected by weights $\mathbf{w}$ which are optimized during training, and biases $\mathbf{b}$, also optimized, are added at every layer of the ANN.
\\
ANNs are universal approximators  \cite{Hornik1991, csaji2001approximation}, meaning that in the limit of an infinite hidden layer they can approximate any continuous function to arbitrary precision.
ANNs have recently been used to predict total energies of MNP \cite{Chiriki2017, Jindal2017, Jindal2018}, study their thermodynamic stability \cite{Chiriki2016}, explore their phase diagram \cite{Chiriki2017}, aid the search for minima structures \cite{Artrith2014a, Kolsbjerg2018, Hajinazar2019}, and run molecular dynamics (MD) simulations \cite{Artrith2013, Artrith2015, Kang2018}.
The price to pay for such flexibility and expressive power is that ANNs are very data-hungry, and typically require orders of magnitude more training points than linear regression or GPR methods to reach the desired accuracy.
In the specific case of an ANN with two hidden layers, the prediction for a local energy $\epsilon(\mathbf{q}_n)$ is given by:
\begin{equation}
\hat{\epsilon}(\mathbf{q}_n) = 
\phi \left\{
b_3 + \sum_{j_{2}} w_{23}^{j_2} 
\phi \left[ b_2^{j_2} + \sum_{j_{1}} w^{{j_1}{j_2}}_{12} 
\phi \left( b^{j_1}_1 + \sum_{{i}} w_{01}^{ij_{1}} \, (\mathbf{q}_n)_i 
\right) \right]  \right\},
\label{eq::neural_network}
\end{equation}
%
%
where $\phi$ is an activation function (sigmoid, hyperbolic tangent, etc.), the $j_l$ indices run over all nodes of layer $l$ (here $l = 1, 2$), the $\mathbf{w}_{l,l'}$'s are the weights connecting the nodes of layer $l$ to the nodes of layer $l'$, $\mathbf{b}_l$ are the biases added to the nodes of layer $l$.
The structure of the artificial neural network just described can be visualised in Figure \ref{fig:regression_comparison} and it is easy to see how Eq. \ref{eq::neural_network} generalises to networks containing more than two hidden layers.
\\

\paragraph{Training}
The training of an ANN consists in the search of weights $\mathbf{w}$ and biases $\mathbf{b}$ which minimize a loss function $L$ on a training set containing $N_{tr}$ points.
Typically the squared error loss (Eq. \ref{eq:loss}) is chosen for regression problems such as energy and/or force fit .
The parameter space $\{ \mathbf{w}, \mathbf{b} \}$ contains from thousands to millions of parameters, and the optimization task is not trivial as the loss function is highly non-convex, presenting many local minima.
For this reason, non-stochastic gradient descent on the loss landscape induced by the full training set would be non-optimal.
Therefore, batch training is typically used to introduce stochasticity into the optimization: a subset of training points is selected and a number of gradient descent steps is taken; this process is then iterated.
Typically, the training process is stopped once the error on the validation set starts increasing, indicating that the neural network has reached the point when it is starting to over-fit on the training data.
\\

\paragraph{Beyond feed-forward artificial neural networks}
While only feed-forward ANNs are covered in this section, a plethora of variations on neural networks have been developed in recent years: recurrent neural networks, convolutional neural networks, variational autoencoders, and generative adversarial networks to name a few.
\cite{Goodfellow2016}
These methods share the same principles of weight and bias optimization via batch training, but contain differences (e.g. in layer structure) which substantially modify the tasks each method excels at.
To the authors' knowledge, there is no existing literature where beyond-feed-forward neural networks are applied to the generation of force fields for MNPs specifically, but they have been employed for other systems. \cite{Schutt2017, Ryczko2018}
\\

\subsection{Gaussian Process Regression}
\label{sec:GPR}
\paragraph{Formalism}
GPR is a Bayesian technique used for supervised learning tasks.
Typically, GPR can require orders of magnitude less data points than ANNs to train, which makes it a suitable choice when data is expensive to generate and/or scarce.
Another advantage of GPR techniques is that, being a fully Bayesian approach to regression, it is possible to estimate an uncertainty associated to every prediction.
\\
As for the two previously described algorithms, here too the local energy function is learnt (instead of the total energy), and then summed according to Eq. \ref{eq::global_energy} to obtain the total energy.
For educational purposes, in the following we assume that a database of local energies pertaining to local atomic environments is available.
This is of course not the case in reality, and the interested reader is referred to \textcite{Bartok:2015iw} and \textcite{Glielmo2019} for more details on how a local energy can be learnt using a database of total energies and forces.
\\
GPR assumes that outputs are distributed like a Gaussian stochastic process.
Given a training data set containing $N_{tr}$ local atomic environment descriptors $\{ \mathbf{q}_i \}$, $i= 1, \dots, N_{tr}$, and a local energy \emph{kernel} function $k(\mathbf{q}_i, \mathbf{q}_j)$, the prediction yielded by the trained GPR on a new local atomic environment descriptor $\mathbf{q}^*$ is:
\begin{equation}
\hat{\epsilon}(\mathbf{q}^*) = \sum_{i=1}^{N_{tr}} k(\mathbf{q}^*, \mathbf{q}_i) \alpha_i
\label{eq::gp_energy_pred}
\end{equation}
where $\alpha_i$ are the weights computed during training via a straightforward matrix inversion.
As already mentioned, we can also compute the uncertainty $\text{VAR}(\mathbf{q}^*)$ associated with a prediction $\hat{\epsilon}(\mathbf{q}^*)$:
\begin{equation}
\text{VAR} (\mathbf{q}^*) = k (\mathbf{q}^*, \mathbf{q}^*) + \sigma_n^2 - \mathbf{k}^{\rm{T}} (\mathbb{K} + \mathbb{I}\sigma_n^2 )^{-1} \mathbf{k}, 
\label{eq::gp_energy_var}
\end{equation}
where $\mathbf{k}$ indicates the vector containing the local energy kernel function evaluated between $\mathbf{q}^*$ and $\mathbf{q}_i$ for $i = 1, \cdots, N_{tr}$, $\mathbb{K}$ is the Gram matrix containing the kernel function evaluated between every pair of inputs $(\mathbf{q}_i, \mathbf{q}_j)$, with $i, j = 1, \cdots, N_{tr}$, and $\sigma_n$ is a hyperparameter that governs the noise associated with the training data.
This uncertainty (Eq. \ref{eq::gp_energy_var}) is such that the error incurred by our GPR when predicting the total energy $\epsilon(\mathbf{q}^*)$ is expected to be normally distributed with mean zero and standard deviation $\sqrt{\text{VAR} (\mathbf{q}^*)}$.
\\
It is possible to learn from atomic forces instead of, or in conjunction with, total energies; this requires the construction of a kernel function for forces.
In order to preserve energy conservation of the force field, such a force kernel function can be obtained by differentiation of an energy kernel function:
\begin{equation}
\mathbf{K}_{FF} (\mathbf{q}_i, \mathbf{q}_j) = \dfrac{\partial^2 k(\mathbf{q}_i, \mathbf{q}_j)}{\partial \mathbf{r}_i \partial \mathbf{r}_j^{\rm{T}}},
\label{eq::force_kernel}
\end{equation}
where $\mathbf{r}_i, \mathbf{r}_j$ indicate the Cartesian coordinates of the central atom in the local atomic environments $i$ and $j$ respectively.
%
\\

\paragraph{The kernel function}
The structure of the kernel function $k$ is of great importance in GPR since its properties directly affect the properties of the force field it generates.
The kernel function acts as a similarity function between pairs of descriptors of local atomic environments $(\mathbf{q}_i, \mathbf{q}_j)$.
Kernel functions also allow for the use of descriptors which do not inherently posses all the required invariance properties, since these can be imposed on the kernel function itself.
For instance, invariance over the rotations can be enforced by performing an Haar integration over the $O(3)$ group \cite{Glielmo2018, Glielmo2017}, invariance over translations can be enforced by integrating over $\mathbb{R}^3$, \cite{Willatt2019} and invariance over permutations can be enforced by direct sum over the permutation group \cite{Glielmo2018, Vandermause2019}.
\\
Some kernels commonly used in literature for ML-FF generation will now be described.
\\

The SOAP energy-energy kernel $k_{SOAP}$ can be straightforwardly built from the power spectrum descriptor derived in Eq. \ref{eq:power_spectrum}: the dot product between normalized power spectrum descriptors of two local atomic environments ($\mathbf{\hat{q}}_{PS}^n$, $\mathbf{\hat{q}}_{PS}^m$) is elevated to a power to obtain:
\begin{equation}
k_{SOAP} (\mathbf{q}^n_{PS}, \mathbf{q}^m_{PS}) = \left( \mathbf{\hat{q}}_{PS}^n \cdot \mathbf{\hat{q}}_{PS}^m \right)^{\zeta}
\label{eq:soap_kernel}
\end{equation}
where $\zeta$ is a positive integer. \cite{Bartok2013}
\\

The $n$-body energy-energy kernel for any finite $n$ can be built from the $n$-body explicit descriptors (Eqs. \ref{eq:2bdescr}, \ref{eq:3bdescr}) by, for example, taking the Gaussian difference of the descriptors summed over the relevant permutation groups and over each dimension of the descriptor.
In the case of the 2-body kernel this results in:
\begin{equation}
k_2(\mathbf{q}^n_2, \mathbf{q}^m_2) = \sum_{i \in \rho_n} \sum_{j \in \rho_m} e^{- \dfrac{(r_{ni} - r_{mj})^2}{2 \sigma ^2}},
\label{eq:2bker}
\end{equation}
where $\sigma$ is the characteristic lengthscale of the kernel.
The 3-body kernel instead reads:
\begin{equation}
k_3(\mathbf{q}^n_3, \mathbf{q}^m_3) = \sum_{i > j \in \rho_n} \sum_{k > l \in \rho_m} \sum_{\mathbf{P} \in \mathcal{P}_c} e^{- \dfrac{\|(r_{ni}, r_{nj}, r_{ij})^{\rm{T}} - \mathbf{P}(r_{mk}, r_{ml}, r_{kl})^{\rm{T}}\|^2}{2 \sigma ^2}},
\label{eq:3bker}
\end{equation}
where $\mathcal{P}_c$ ($ |\mathcal{P}_c | = 6$) is the set of permutations of three elements.
\\

\paragraph{Mapping force fields}
One of the main drawbacks of GPR FFs is that, despite being computationally faster than \emph{ab-initio} methods to evaluate, they scale linearly with the number of training points, $N_{tr}$ (see Eq. \ref{eq::gp_energy_pred}).
This burden can be avoided for a particular sub-class of kernel functions, and the computational cost of the so-derived mapped force fields (MFFs) is comparable to classical tabulated force fields. \cite{Zeni2018}
The constraint that must be imposed on the kernel function in order for the FF to be mappable is that its order must be finite \cite{Glielmo2018} or, in other words, that the kernel function is not many-body.
\\
The mapping procedure starts with the identification of the maximum order of the kernel function used, for example 2-body for kernels which only depend on interatomic distances from central atoms, 3-body for kernels that use angles, 4-body for kernels that use torsion angles, and so on.
Subsequently, the local energy prediction made by the GP is calculated and stored on an array of values.
For example, for a 2-body kernel, the bond energy can be calculated and stored on an array of distances between 1 $\text{\AA}$ and the cutoff radius $r_c$.
This process yields a tabulated force field of the same order of the kernel.
Predictions during MD runs can then be obtained by using an interpolator on the stored values for every pair, triplet, or quadruplet (when using, respectively, 2, 3 or 4-body kernels) of atoms in the system.
This method has been shown to increase computational speed by a factor $10^4$ with respect to traditional GPR without incurring in any additional accuracy loss. \cite{Glielmo2018, Zeni2018}
The interested reader is directed to MFF, a Python package that implements the above method within the framework of GPR FFs: \url{https://github.com/kcl-tscm/mff}. \cite{mff_package}

\section{Database selection}
\label{sec:database}
\begin{figure}[h!]
\begin{center}
\includegraphics[width=0.5\columnwidth]{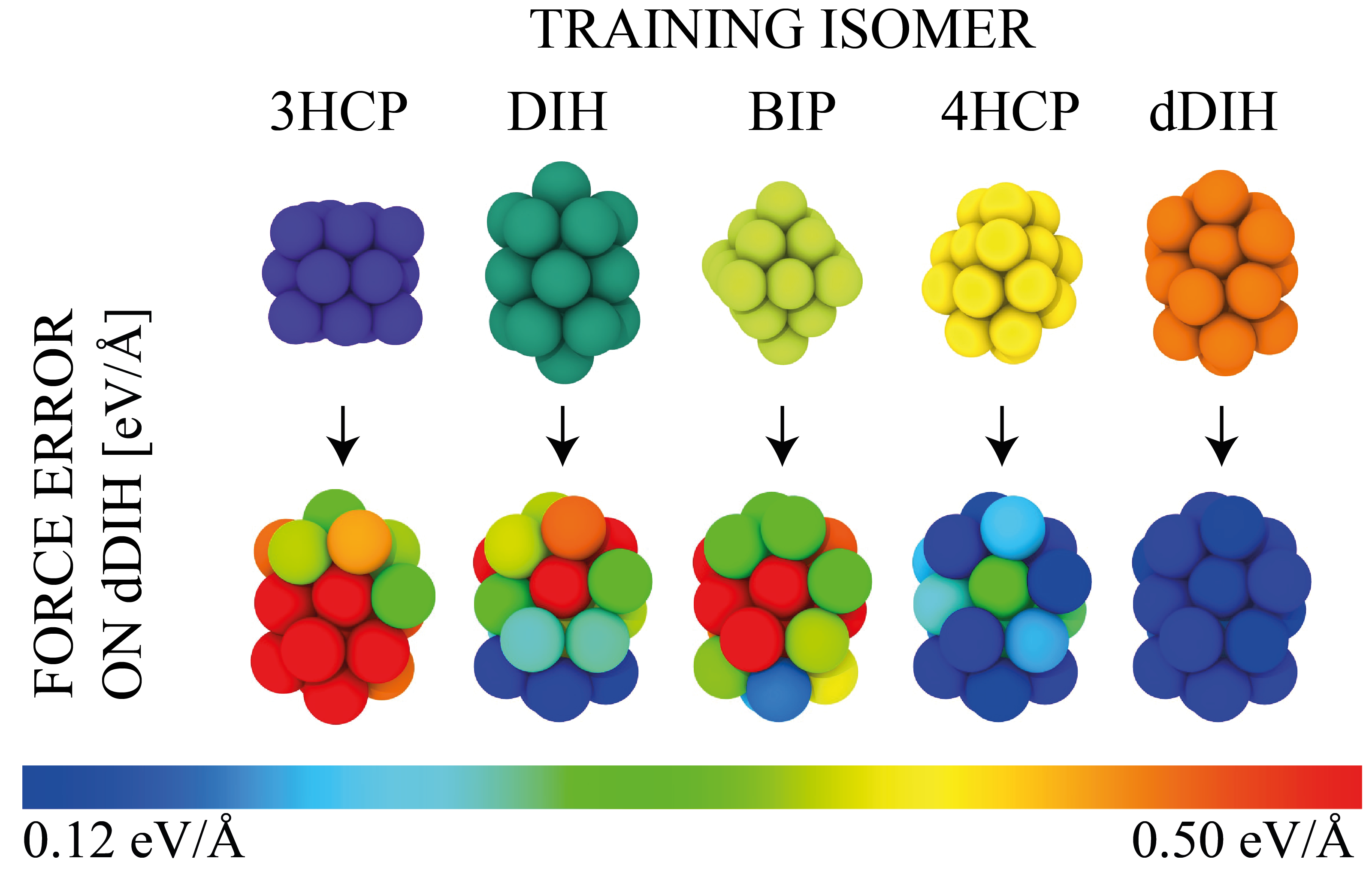}
\caption{Mean absolute error on the DFT force vector incurred by a 3-body GPR algorithm trained on five isomers of Ni$_{19}$ and tested on the defected double icosahedron (dDIH) isomer.
Figure adapted with permission from \textcite{Zeni2018}.}
\label{fig:ni19-cross}
\end{center}
\end{figure}

\begin{figure}[h!]
\begin{center}
\includegraphics[width=0.4\columnwidth]{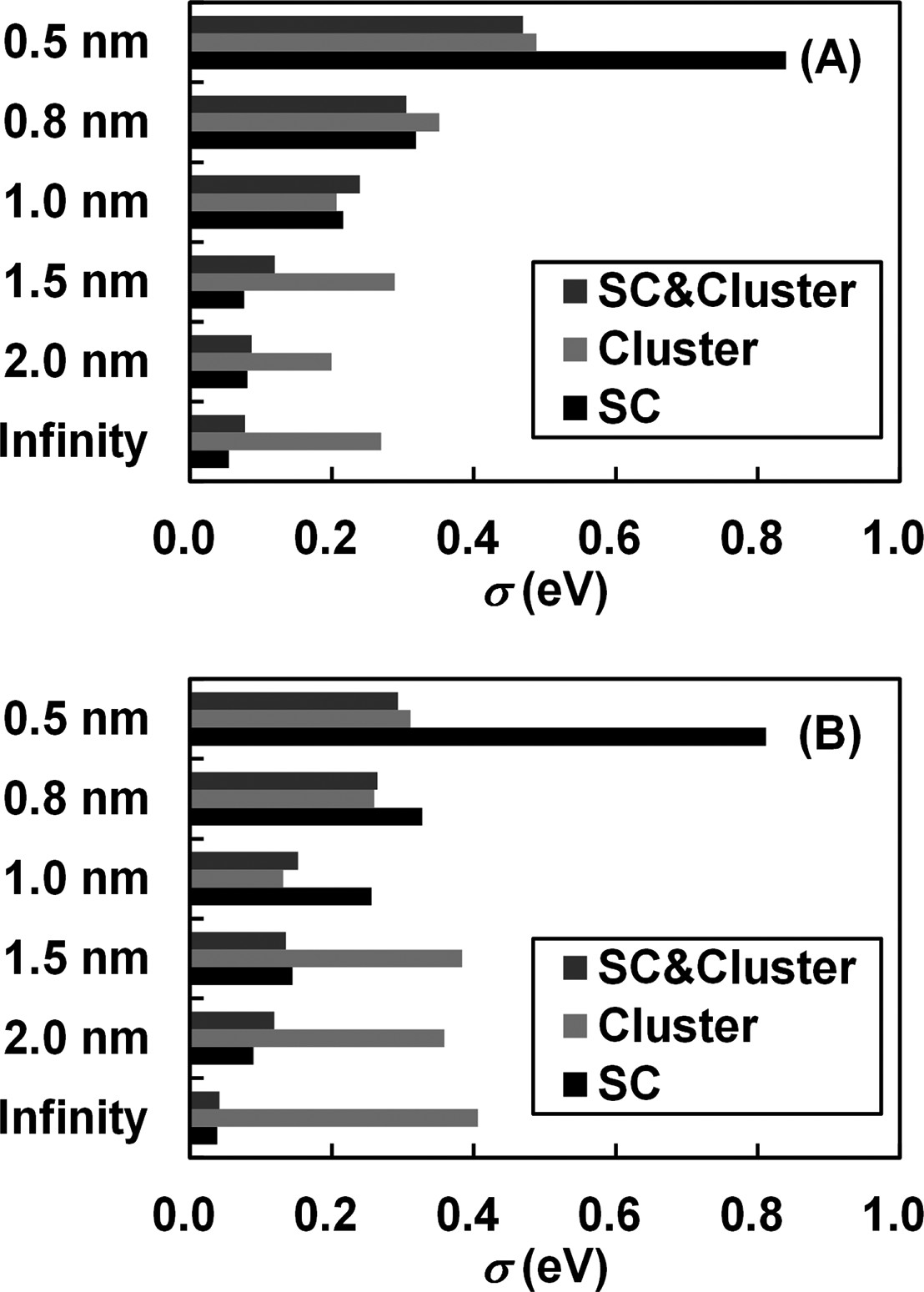}
\caption{Mean absolute errors on the binding energies of O; the Rh$_{1-x}$Au$_x$ single crystals and particles are predicted using DFT data of single-crystal surfaces, small clusters, or both; using cutoff radii of (A) 6 and (B) 8 $\text{\AA}$.
Figure reprinted with permission from \textcite{Jinnouchi2017a}.
Copyright (2017) American Chemical Society.}
\label{fig:jinnouch2017}
\end{center}
\end{figure}

The accuracy of a ML-FF depends on whether the predictions are made in an extrapolation or interpolation region.
Without giving a formal definition, it can be helpful to think of an interpolation region as that region of the  descriptor space which is ``close" to the training database.
Generally speaking, accuracy is higher, and predicted variance (in the case of GPR) is lower, when predictions are made in an interpolation regime.
Therefore, it is preferable to always work in interpolation and, if possible, detect when and where the algorithm is extrapolating.
Enforcing the force field to work in an interpolation regime becomes crucial in the case of nanoparticles and nanoalloys, where the number of competing isomers, even at small sizes, is very large. 
Due to this inherent complexity of the configuration landscape, it would be desirable to include in the initial training set a diverse array of isomers which yield accurate predictions also on never-seen-before geometries and/or chemical orderings.
Two example cases extracted from the literature may help set this problem into practical scenarios.
\\
In a study on Ni$_{19}$, it was shown that force prediction error was substantially lower if training and testing on the same morphology rather than training and testing on two different MNP morphologies. \cite{Zeni2018}
Furthermore, training on structurally heterogeneous databases was shown to lead to a balanced trade-off between versatility and overall accuracy. 
By the same token, training on low-symmetry or defected structures, which present very different local atomic environments, resulted in a more accurate machine learning model.
This is caused by the higher variety of local atomic environments present is low-symmetry morphologies, which allow the convex hull spanned by the training inputs to encompass a larger phase space.
\\
In figure \ref{fig:ni19-cross} it can be seen how training on low-symmetry structures (4HCP) yields force fields that are more accurate on a target morphology w.r.t. training on high-symmetry structures (3HCP, DIH, BIP).
\\
In the case of a work focusing on AuRh nanoparticles, it was also reported that regression models predict binding energies of N, O, C molecules correctly or not depending on the size of the nanoparticles and surfaces in the training set.  \cite{Jinnouchi2017a}
Figure \ref{fig:jinnouch2017} shows how training sets which contained only single crystal training points incur in high errors when predicting binding energies on nanoparticles, and viceversa.
On the other hand, training sets which contain both the nanoparticle and the single crystal morphologies consistently incur in low binding energy mean absolute errors for MNPs ranging from diameter 0.5 nm to bulk.
\\
\\
A diverse set of strategies has therefore been developed in the machine learning community to build non-biased and automatically comprehensive training databases.
These databases must be such that the machine learning force field generates predictions in an interpolation regime for the study of interest, and must assure that redundant, costly, information is not gathered for small regions of the phase space with respect to the investigation of interest.
\\
Training database selection algorithms can be broadly classified into two categories: either the database needs to be generated ad-hoc (active learning), or it must be sub-sampled from a larger, already existent database (sparsification).
In the following sections we introduce the aforementioned concepts and extract some examples from literature to better inform the reader.

\subsection{Active learning} 
\label{sec:active_learining}
If there is no sufficient pre-existent \emph{ab-initio} training database for the task of interest, one convenient way to minimize the computational effort in building it is active learning.
Active learning indicates a framework where the ML algorithm is iteratively trained on structures where the algorithm's predictions are deemed uncertain. 
If this is the case, forces and energies are computed with \emph{ab-initio} methods on the structures where the ML predictions were not accurate enough, and finally inserted into the training database.
While GPR can naturally return a predicted variance (Eq. \ref{eq::gp_energy_var}), in the case of ANNs an uncertainty estimate can be obtained as the variance of the predictions made by multiple ANNs trained for the same task.  \cite{Artrith2012, gastegger2018molecular}
\\
An example use of this approach can be found in conjunction with a genetic algorithm and a basin hopping scheme in \textcite{Jennings2019}.
Here, ML-FFs are used as a computationally inexpensive energy predictors so to fast-screen for energetically relevant isomers and thus facilitating the quick convergence of the optimization algorithms driving the energetic landscape exploration as well as the surrogate machine learning model training.  
\\
In a similar fashion, surrogate machine learning force fields can be also used to first quickly explore energetically relevant portions of the conformational space and rearrangement pathways via molecular dynamics and gather diverse ensembles of nanoparticle structures for which electronic structure calculations are executed. 
These are then fed as novel energy and force training points to the machine learning model in an iterative fashion until convergence of the model in the region of interest. 
 \cite{Kolsbjerg2018,Ulissi2017}

\subsection{Database sparsification}
\label{sec:sparsification}
If a database which is considered comprehensive for the system of interest is available, it can be useful to sub-sample a smaller training set from the full dataset, in order to reduce the computational cost of training and using a ML algorithm.
This is especially beneficial in the case of GPR, where the training computational cost scales as $\mathcal{O}(N_{tr}^3)$ and the prediction cost scales with $\mathcal{O}(N_{tr})$, where $N_{tr}$ is the number of training points.
Many database sparsification methods exist, but up to now only few applications relative to the construction of ML-FFs can be found in literature.
Nonetheless, we will briefly present some database sparsification methods for ML-FFs, namely: farthest point sampling, measured error sampling, CUR decomposition, and descriptor-space sampling.
\\

\paragraph{Farthest point sampling}
Farthest point sampling (FPS) is a database selection method where points are iteratively added to a training set based on their distance, according to a pre-selected metric, from the points already present in the training set.
At each iteration, distances between the points inside and the points outside of the training set are computed, and the point furthest away from the training set, according to the chosen distance metric, is included.
The computational cost of this method is $\mathcal{O}(N_{tr} \cdot N)$, where $N$ is the total size of the database to sub-sample from.
FPS sampling has been employed in literature to reduce training set size for Gaussian Process regression methods. \cite{Bartok2017, Musil2018}
While using GPR, predicted variance can be used instead of distance from training set as a way to measure diversity.
In this framework, points are iteratively added to the training set based on the value yielded by Eq. \ref{eq::gp_energy_var}.
\\

\paragraph{Measured error sampling}
Another class of iterative methods, which can be used independently of the ML framework employed, is based on the measured error incurred by the ML algorithm.
The starting training database contains a small number of randomly sampled data points, and is then progressively expanded by inclusion of data points on which the measured incurred error is maximum.
This method should be effective in reducing the maximum error incurred on the global database at the cost of an increased computational complexity w.r.t. other similar methods, e.g. FPS.
\\

\paragraph{CUR decomposition}
CUR decomposition is a matrix-approximation technique that decomposes a matrix $\mathbf{A}$ into a product of three matrices $\mathbf{C}$, $\mathbf{U}$, $\mathbf{R}$. \cite{mahoney2009cur}
The method is similar to a low-rank single value decomposition approximation, but is built so that $\mathbf{C}$ is composed of columns of the original $\mathbf{A}$, and $\mathbf{R}$ from rows of the same $\mathbf{A}$.
CUR decomposition can be applied to GPR by first building the Gram matrix $\mathbb{K}$ between all the points in the full dataset, and then reducing its rank by selecting the $N$ columns (or rows) that compose $\mathbf{C}$ (or $\mathbf{R}$).
The method has a computational scaling of $\mathcal{O}(N^2)$; this can be reduced by subdividing the entire datasets into batches of size $N_{batch} \ll N$, thus introducing an approximation but reducing the scaling to $\mathcal{O}(N_{batch}^2 \cdot B)$ where $B$ is the number of batches.
Note that this method is not only helpful in identifying a sparse but information-rich database, but also in determining non redundant sets of descriptors. \cite{Imbalzano2018}
\\

\paragraph{Descriptor-space sampling}
\begin{figure}[!tbp]
  \centering
  \begin{minipage}[b]{0.48\textwidth}
    \includegraphics[width=\textwidth]{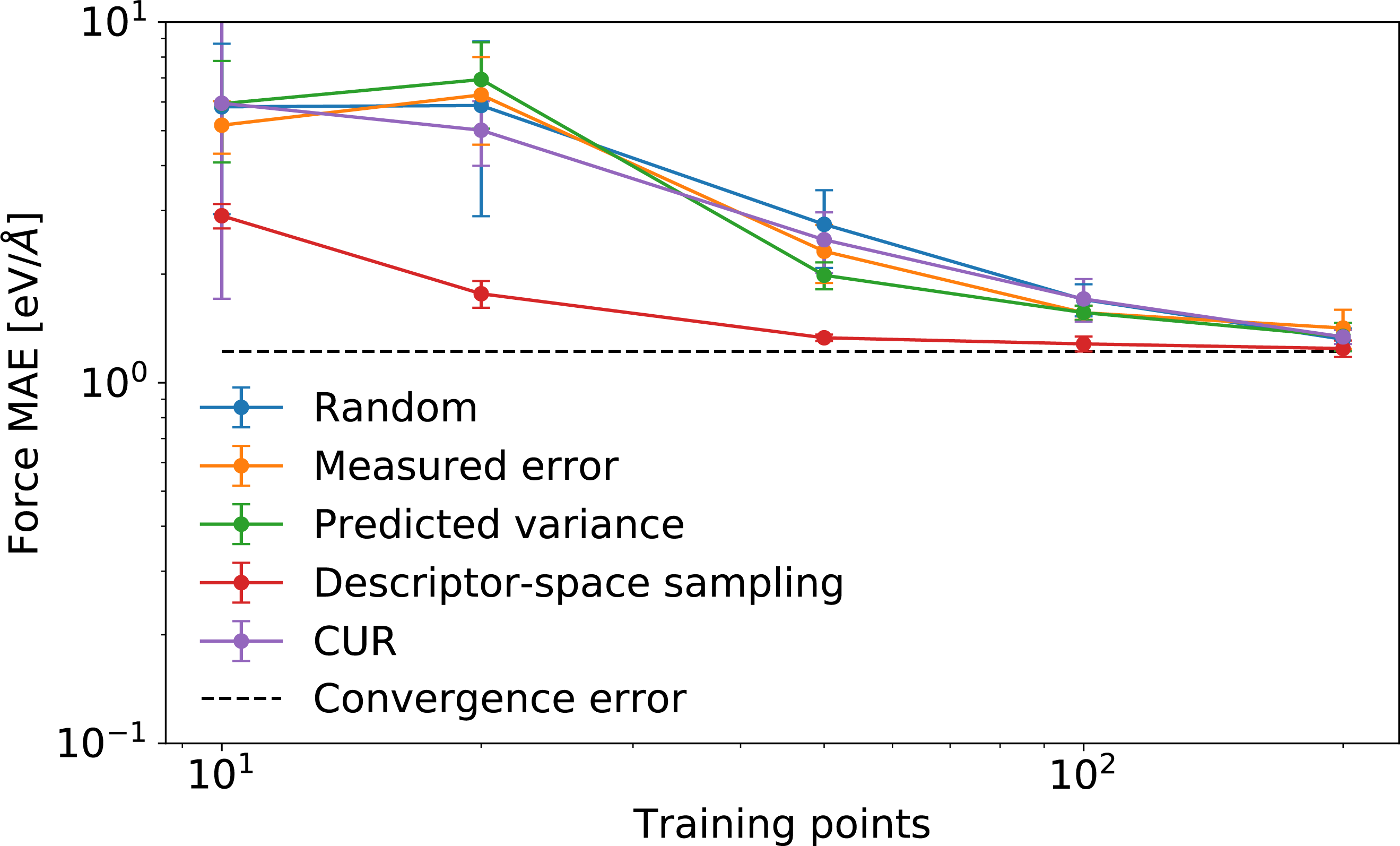}
  \end{minipage}
  \hfill
  \begin{minipage}[b]{0.48\textwidth}
    \includegraphics[width=\textwidth]{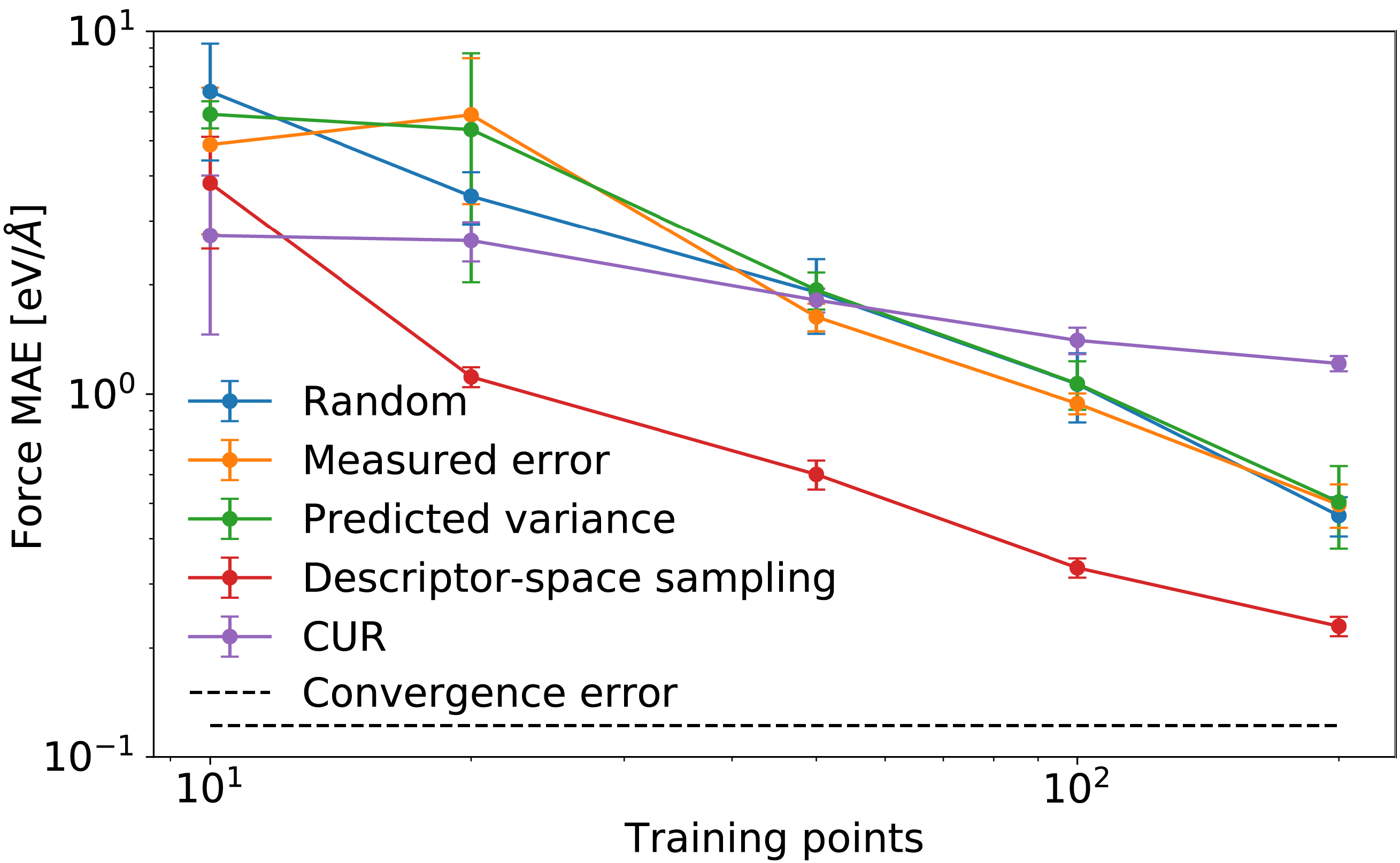}
  \end{minipage}
\caption{Log mean absolute error on force vectors incurred by a 2-body (left) and a 3-body (right) GPR algorithm as a function of the log number of training points $N_{tr}$ when trained and tested on a total database of (PtCu)$_{13}$ local atomic environments, with $N$ = 13000.
The training points have been selected according to the three algorithms described above, and also with random sampling. 
The black dashed line indicates an estimate of the convergence error, obtained by training the algorithm on 2000 randomly selected training points.
The error bars display the standard deviation of the error, obtained by repeating the process five times.
}
\label{fig:PtCu_sampling}\end{figure}
The following sampling method is inspired by the database sparsification algorithm used in \textcite{Bernstein2019}, but more general in its formulation.
First, a physical descriptor must be chosen e.g. distance from, or angle w.r.t., the central atom of local atomic environments.
Secondly, a set of discrete bins spanning the possible values the chosen descriptor can take must be built (e.g. a 1-D histogram containing distances from 1.5 $\text{\AA}$ to 5 $\text{\AA}$ binned every 0.05 $\text{\AA}$).
Local atomic environments in the total database are then shuffled and accepted into the training set if they contain at least one occurrence of a binned descriptor's value which is not already present in the training data set.
This simple method assures that the final training set contains environments which are as diverse at possible in the descriptor's space.
This method is very computationally efficient, as only a single pass over the database is required.
\\

Fig. \ref{fig:PtCu_sampling} shows a comparison of the mean absolute error on force vectors of three of the aforementioned sampling methods when building a ML-FF using GPR on a database containing a total of 13000 (PtCu)$_{13}$ structures, it is evident how descriptor-space sampling is the best performing algorithm for this system.
For tests done with the measured error, predicted variance and CUR database selection methods, the total database was subdivided into batches and training sets were built by grouping the points selected in each of these batches.
This was done in order to reduce the computational cost of such methods.
It is interesting to note how, for this particular system, GPR algorithms trained using databases selected by methods other than descriptor-space sampling incur in errors similar to the ones yielded by randomly chosen training datasets.
Such poor performance could be caused by the batch approximation used.
In any case, the descriptor-space sampling method has a better scaling than any other method presented, is overall faster to compute even on small datasets, and shows a good performance on the analysed dataset.
\\

\section{Applications of ML-FFs for metallic nanoparticles}
\label{sec:examples}

After having introduced the formalisms behind the machine learning algorithms for generating fast and accurate energy predictions, we focus on their applications to understand structural properties of metallic nanoparticles.
This section is divided into two parts. 
The first subsection reports applications where ML-FFs were used to supplement or substitute \emph{ab initio} calculations for energy calculations, e.g. for global minima  structure search or estimations and mapping of activation energies.
The second subsection focuses on applications of ML-FFs to finite temperature simulations over several tens of ns, carried out e.g. to estimate phase diagrams and/or thermal properties of metallic nanoparticles.

\subsection{ML-FFs for energy calculations}
\label{sec:ml4gms}
Machine learning force fields can be used as a surrogate for DFT when exploring the configurational space of nanoparticles.
These ML-FFs must exhibit good accuracy especially when predicting energies, and can be trained either \emph{a priori} or on-the-fly, depending on the size of the available database.
In this section, we briefly resume the state-of-the-art on energy prediction yielded by ML-FFs to sample local minima, including chemical re-ordering, and to estimate adsorption energies.
\\

\paragraph{Monometallic nanoparticles}
\begin{figure}[h!]
\begin{center}
\includegraphics[width=0.80\columnwidth]{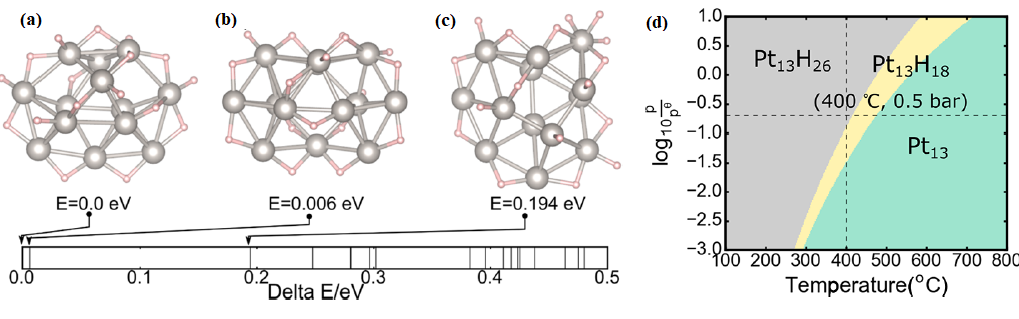}
\caption{\textbf{(a)}, \textbf{(b)}, \textbf{(c)} Relative energies (eV) of the 20 Pt$_{13}$H$_{18}$ isomers in the low energy metastable ensemble (zero represents the energy of the global minimum) and
structures of the three most stable ones. 
Grey spheres indicate Pt atoms while red ones indicate H atoms.
\textbf{(d)} Thermodynamics stability of Pt$_{13}$H$x$ clusters (x = 0, 18, 26)
as a function of temperature and hydrogen pressure.
Figure adapted with permission from \textcite{Sun2018}.
Copyright (2018) American Chemical Society.}
\label{fig:sautet2018}
\end{center}
\end{figure}
\begin{figure}[t!]
\begin{center}
\includegraphics[width=0.95\columnwidth]{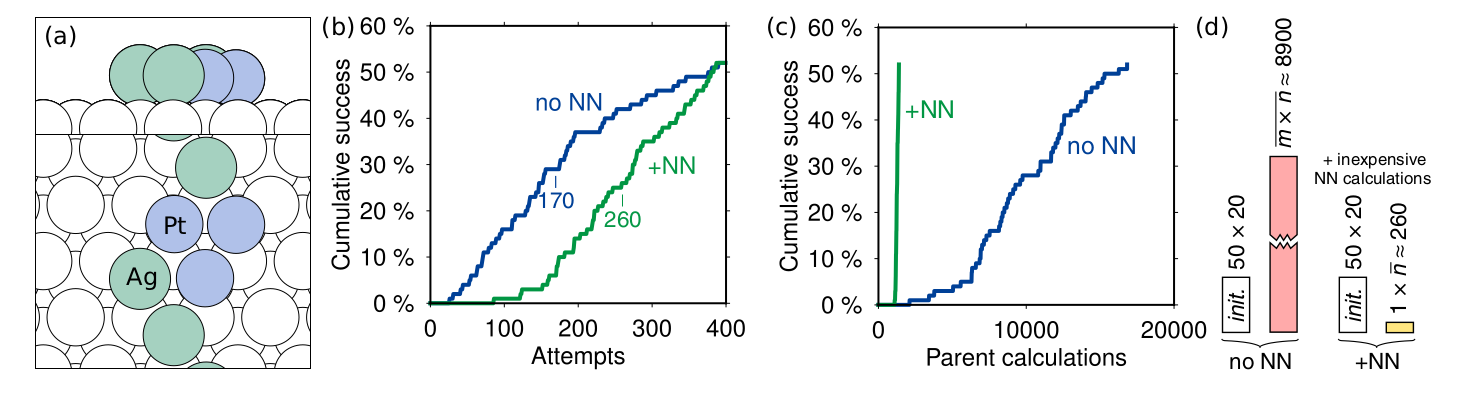}
\caption{\textbf{(a)} The atomistic model of a Pt$_{3}$Ag$_{3}$ nanoparticle supported on a Pt(111). Ag and Pt atoms colored green and blue, respectively, the slab Pt atoms are white.
 \textbf{(b)} \textbf{(c)} The success rate of locating the global minimum as a function of the number of candidates evaluated and the number of needed parent calculations, respectively.
\textbf{(d)} Graph highlighting the significant reduction in the average parent calculations needed.
Figure reprinted with permission from \textcite{Kolsbjerg2018}.
Copyright 2018 by the
American Physical Society .}
\label{fig:hammer2018}
\end{center}
\end{figure}
The first application of machine learning to speed up the sampling of monometallic nanoparticles' energetic landscape consisted in the use of an ANN employing ASFs (Eqs. \ref{eq:ASF2}, \ref{eq:ASF}) as descriptors to search for global minima of gold nanoclusters within the basin-hopping method by Ouyang, Xie and Jiang.
A new putative global minimum which has a core-shell structure of Au$_{10}$@Au$_{48}$ and C4 symmetry was found, highlighting the benefit of comprehensive and fast exploration powered by neural network force fields.\cite{Ouyang2015}
\\
Subsequently, Zhai and Alexandrova employed a GPU accelerated ANN for sampling the potential energy surface of Pt$_{9}$ and Pt$_{13}$. 
The authors used a 4-body descriptor as input to the ANN, and carried out global minima searches where the ML-FF was used to complement the DFT calculations and speed up the process.
Finite temperature effects were also included a posteriori to probe the population distributions of such systems. 
This comprehensive study showed that the ensemble-averaged vertical ionization potential of the sytems under investigation changes when temperature increases, and that the catalytc property under \emph{operando} conditions can be different from that evaluated at the global minimum structure. \cite{Zhai2016}
\\
More recently, Sun and Sautet presented an application of a genetic algorithm coupled with a high dimensional neural network potential using ASFs as descriptors to accelerate the comprehensive search for low energy metastable structures of Pt$_{13}$ under different H pressures. 
The presence and influence of these Pt$_{13}$ structures during catalysis was discussed for hydrogen evolution reaction and methane activation. 
Although the ensemble of accessible metastable structures is relatively small under reaction condition, these structures can strongly influence the experimentally observed activity.\cite{Sun2018}
Figure \ref{fig:sautet2018} reports the lowest energy isomers found for Pt$_{13}$ in the presence of H, together with the thermodynamic stability diagram of Pt$_{13}$H$_x$ clusters.
\\
The same year, Kolsbjerg et al. looked at putative global minimum structures for Pt$_{13}$ on a MgO support.
The authors used an on-the-fly trained ANN force field to relax structures to local minima during evolutionary algorithm searches; here, ASFs were the descriptors of choice.
The computational speed-up inherent to the framework most importantly enabled the screening of hundreds of kinetic rearrangement pathways connecting different low-energy conformers.\cite{Kolsbjerg2018}
Figure \ref{fig:hammer2018} is shown to highlight the decrease in the number of \emph{ab-initio} calculations performed thanks to the use of an ANN during the minima structure search. 
\\

\begin{figure}[h!]
\begin{center}
\includegraphics[width=0.80\columnwidth]{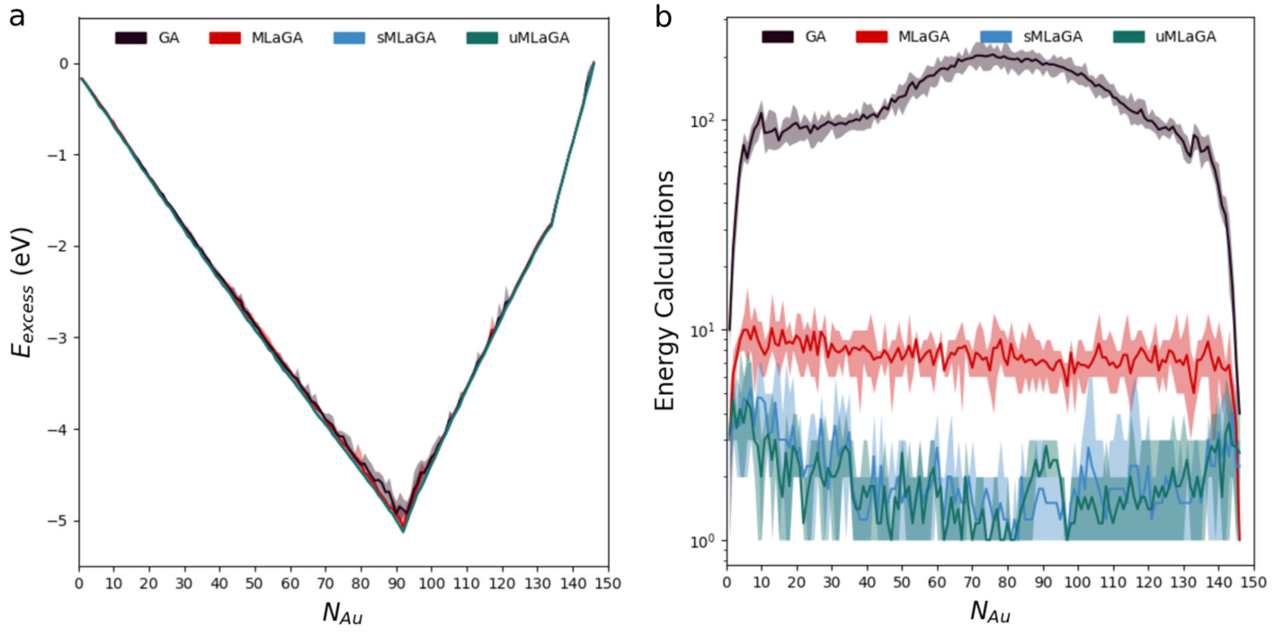}
\caption{\textbf{(a)} Excess energy of a (PtAu)$_{147}$ nanoparticle as a function of its chemical composition, located with a ML-accelerated genetic algorithm employing effective-medium theory calculations.
\textbf{(b)} Number of energy calculations as a function of the nanoparticle's composition.
The four lines correspond to traditional genetic algorithms (GA), machine learning accelerated GA (MLaGA), serialized MLaGA and MLaGA utilising uncertainty (uMLaGA); average and standard deviation over five searches are shown.
Figure reprinted with permission from \textcite{Jennings2019}.}
\label{fig:jenning}
\end{center}
\end{figure}

\paragraph{Nanoalloys}
The computational speed-up offered by ML-FFs to global minima search is key in the study of nanoalloys, as the extra degree of freedom given by the number of possible homotops increases the dimensionality of the space that has to be explored.
\\
Jennings and co-workers employed GPR force fields using a built-for-purpose local atomic environment descriptor to search for stable, compositionally variant, geometrically similar PtAu nanoalloys. 
The machine learning approach yielded a 50-fold reduction in the number of required energy calculations compared to a traditional “brute force” genetic algorithm. \cite{Jennings2019}
Figure \ref{fig:jenning} shows how the ML-based genetic algorithm was able to faithfully reproduce the convex hull for the excess energy of a (PtAu)$_{147}$ nanoparticle (left) while reducing by orders of magnitude the number of energy calculations required (right).
\\
One of the latest applications of machine learning force fields for nanoalloys involved the study of a trimetallic system, Cu--Pd--Ag.
The fast and accurate ANN force field, powered by a stratified training scheme, was used to define a force field which, together with multi-tribe evolutionary searches, improved the efficiency in identifying stable elemental (30--80 atoms), binary (50, 55, and 80 atoms), and ternary (50, 55, and 80 atoms) Cu--Pd--Ag clusters. 
The best candidate structures identified with the neural network model, which used ASFs as local atomic environment descriptor, were showcased to have consistently lower energy at the density functional theory level compared with those found with searches employing an initial layer of inter-atomic potentials search.\cite{Hajinazar2019}
\\

\paragraph{Nanocatalysts}
\begin{figure}[h!]
\begin{center}
\includegraphics[width=0.7\columnwidth]{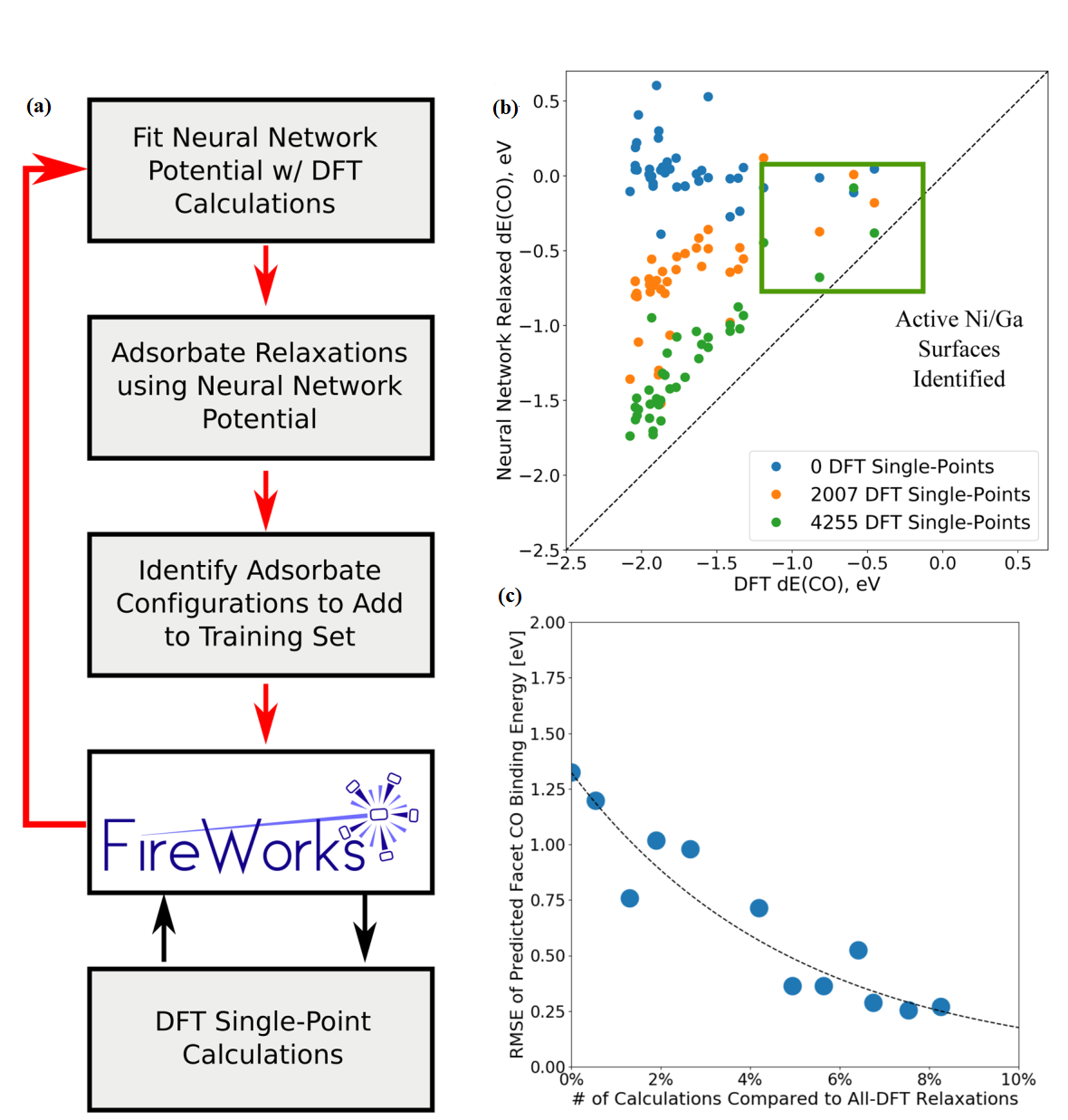}
\caption{\textbf{a} Scheme used for training and use of the model.
New training data is acquired via DFT single-point calculations. 
\textbf{b} Scatter plot for three iterations of the convergence system,
starting from very poor predictions and converging to more accurate predictions of adsorption energy.
\textbf{c} Convergence of the accuracy of the CO adsorption energies with respect to the training set size.
Points possess some inherent noise due to the stochastic nature of the neural network training algorithms.
Figure adapted with permission from \textcite{Ulissi2017}.
Copyright (2017) American Chemical Society.
}
\label{fig:ulissi2017}
\end{center}
\end{figure}
Machine learning energy predictions not only aid the search for the ensemble of energetically relevant nanoparticles' isomers, but also greatly enhance the scope for high throughput probing of active sites properties for catalytic reactions.
\\
One of the first applications of machine learning algorithms to speed up the high throughput characterization of the adsoprtion properties of available sites in a nanoparticle was developed by Ulissi et al..
Active sites for every stable low-index facet of a NiGa bimetallic crystal were enumerated and catalogued while the activity of these sites with respect to CO$_{2}$ adsorption was explored using a neural network-based surrogate model.
This approach, which used ASFs as the local atomic environment descriptor of choice, reduced the number of explicit DFT calculations required for  activity estimates by an order of magnitude.
While most facets had similar activity to Ni surfaces, a few exposed Ni sites showed a very favorable on-top CO configuration. \cite{Ulissi2017}
Figure \ref{fig:ulissi2017} reports the training scheme adopted in their work, along with a scatter plot displaying the increasing accuracy of the ANN model for the CO adsorption energies.
\\
Jinnouchi et al. used linear regression in conjunction with the SOAP similarity function (Eq. \ref{eq:soap_kernel}) to interrogate catalytic activities for the direct NO decomposition on RhAu alloy nanoparticles.
The employed method efficiently and accurately predicted the energetics of catalytic reactions on nanoparticles while providing information on structure-property relationships when combined with kinetic analysis. \cite{Jinnouchi2017a}
\\
The same authors later predicted the binding energies of N, O, and NO with RhAu surfaces and particles using the same approach.
Kinetic analyses of the direct decomposition of NO on RhAu nanoparticles were carried on to demonstrate that catalytic activity increases with a decrease in the particle diameter to 2.0 nm. 
Below a diameter of 1.5 nm a drop in the catalytic activity is registered and rationalized in terms of the disappearance of active alloyed corner sites on the small nanoparticles. 
 \cite{Jinnouchi2017b}
 \\
AuCu nanoalloys' adsorption properties have also been investigated, as reported by J{\"a}ger et al..
Potential energy scans of hydrogen on AuCu clusters and on MoS$_{2}$ surfaces were conducted to compare and assess the accuracy of the Smooth Overlap of Atomic Positions, Many-Body Tensor Representation \cite{huo2017unified}, Couloumb matrix \cite{Rupp2012} and Atom-Centered Symmetry Functions in a kernel ridge regression framework.
\cite{Jager2018} 
\subsection{ML-FFs for finite temperature simulations}
\label{sec:supported}
Beyond sampling local minima in the energetic landscape of MNPs, MD-based studies of kinetic rearrangements and thermodynamic stability of MNPs are also of great interest. \cite{Baletto2005}
With this in mind, employing ML-FFs to run fast MD simulations becomes more and more enticing.
At the nanoscale, this allows to predict the dynamical and thermodynamical properties of MNPs.
In this section we discuss the state-of-the-art of ML-FFs for MD simulations of MNPs and nanoalloys.
\\

\begin{figure}[h!]
\begin{center}
\includegraphics[width=0.80\columnwidth]{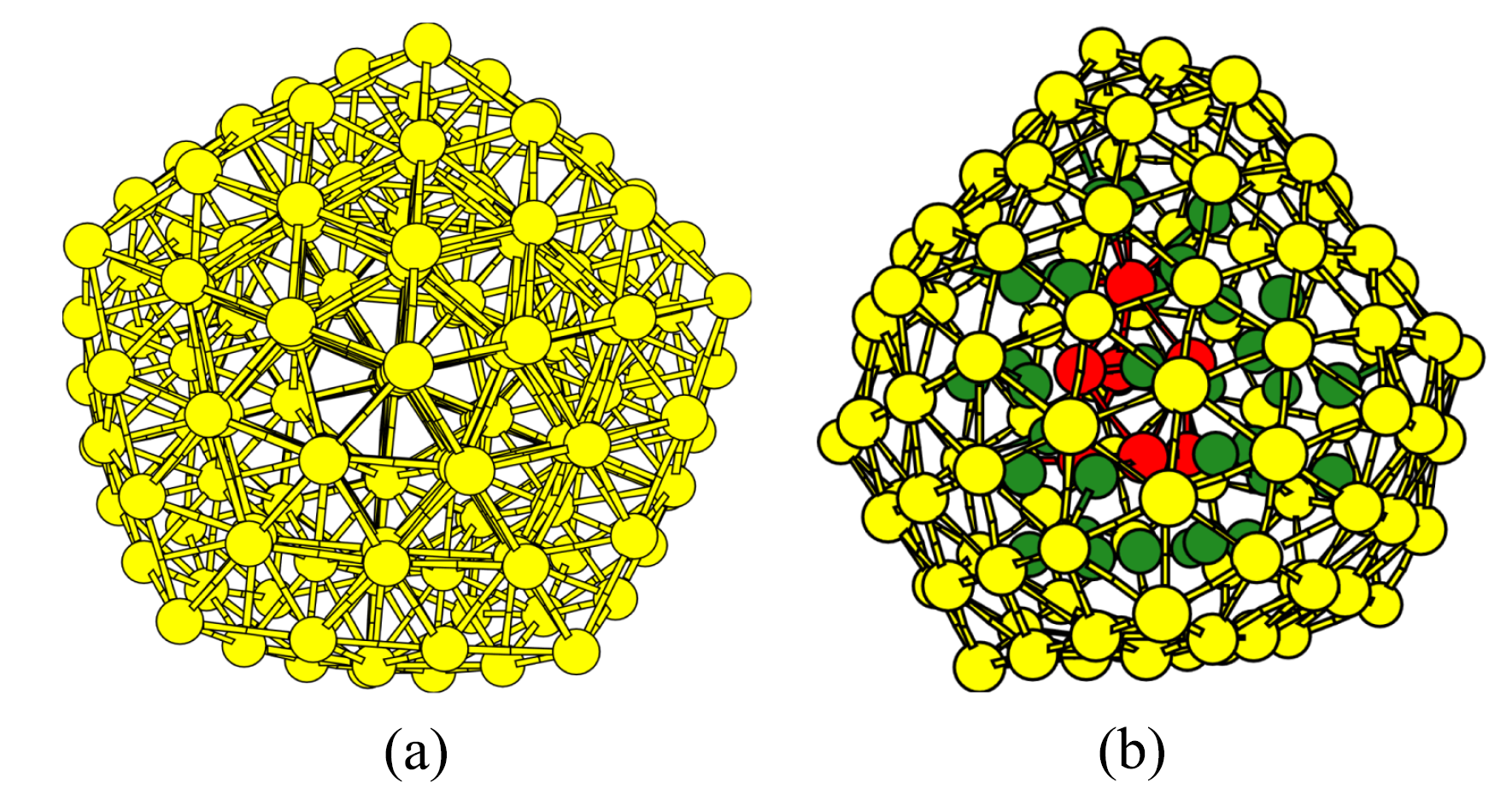}
\caption{\textbf{(a)} Geometry of an Au$_{147}$ icosahedron. 
\textbf{(b)} Geometry of the putative ground minimum for Au$_{147}$ found using an ANN force field.
Figure \textbf{(b)} is color coded so to highlight the three shells of Au atoms.
Figure Reprinted from \textcite{Jindal2017}, with the permission of AIP Publishing.}
\label{fig:jindal}
\end{center}
\end{figure}

\paragraph{Monometallic nanoparticles}
\begin{figure}[b!]
\begin{center}
\includegraphics[width=0.80\columnwidth]{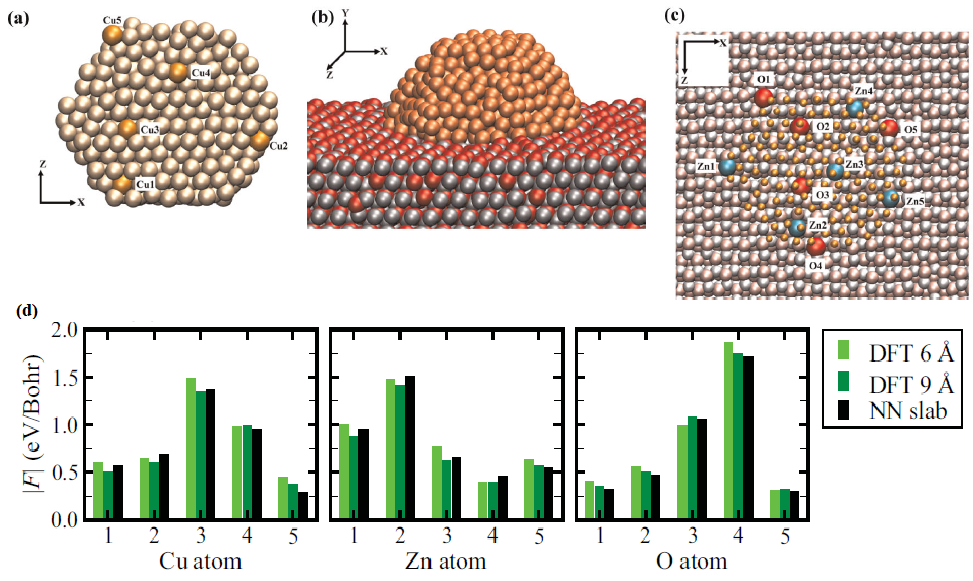}
\caption{\textbf{(a)} Bottom-view of a snapshot of a MD simulation at 1000~K of a Cu$_{612}$ cluster at the ZnO(1010) surface; five copper atoms have been selected to compare the ANN-predicted forces with the DFT forces (in \textbf{d}).
\textbf{(b)} Side view of the cluster.
 \textbf{(c)} A top view of the ZnO(1010) surface is shown.
  Five oxygen and five zinc atoms have been chosen for a closer investigation of the forces (in \textbf{(d)}).
   \textbf{(d)} Comparison of the force modulus of two DFT force evaluations using atoms within 6 $\text{\AA}$ and 9 $\text{\AA}$ from the central one, and the neural network force field on the whole slab for the atoms highlighted in \textbf{(a)} and \textbf{(c)}.
Figure adapted with permission from \textcite{Artrith2013}.
Copyright 2013 WILEY-VCH Verlag GmbH $\&$ Co. KGaA, Weinheim. }
\label{fig:nonguch2016}
\end{center}
\end{figure}
Artrith et al. investigated the structural and energetic properties of copper clusters when supported on zinc oxide in the first ever application of neural network force fields for metallic nanoparticles.
In their seminal work, the authors assess the accuracy of ANN potentials employing the ASF descriptors.
The manuscript builds on top of a previous paper by Artrith and Behler where a defected Cu surface was used as a benchmark for ANN force fields \cite{Artrith2012} providing an estimation of the accuracy and speed possible when using ML-FFs for metallic nanoparticles.
The training of heterogenous ensembles of structures was demonstrated to be transferable to complex large-scale simulations of several defected stepped surfaces and nanoparticles, such as the one displayed in Figure \ref{fig:nonguch2016}. \cite{Artrith2013}
\\
ANN force fields developed in the group of Bulusu, which used the ASF descriptor, were employed to sample local and global minima in the potential energy surface of Na nanoparticles of size 16--40, where transitions were also thoroughly probed using a Monte Carlo scheme.
The accuracy of the force field, and the timescale probed, allowed to establish the presence of a characteristic premelting peak in the heat capacity curve, preciding a main melting peak, for clusters in the 20–-40 atoms size range, \cite{Chiriki2016}, corroborating the observations of stepwise melting in small Na clusters first reported by Aguado and coworkers \cite{Aguado2011}.
\\
The same group later studied Au nanoclusters' energetic and thermodynamic properties.
ANN force fields were employed to probe the potential energy landscape and thermodyanmical behaviour of Au$_{17}$, Au$_{34}$, Au$_{58}$.
Here too, ASFs were the local atomic descriptors of choice to generate inputs for the ANN.
Canonical and microcanonical molecular dynamics sampling was performed for a total simulation time of around 3~ns for each nanoparticle.
The study used such data to demonstrate the presence of a dynamical coexistence of solid-like and liquid-like phases near melting transition.
The investigation further encompassed the estimation of the probability at finite temperatures for a set of isomers lying less than 0.5 eV from the global minimum structure.
For Au$_{34}$, in particular, the global minimum structure resulted far from being the most dominant structure, even at low temperatures. \cite{Chiriki2017}
\\
Later, ANN force fields trained by first-principles density functional theory total energies were applied to search for global minima of gold nanoclusters within the basin-hopping method.
In this case, the same authors decided to employ a descriptor based on the spherical harmonics expansion (Eq. \ref{eq:sh_descr}), reporting its increased efficiency w.r.t. ASFs.
A study on the fluxionality in Au$_{147}$ was performed, and it was concluded that the system presents a dynamic surface. 
Such observation was concluded to be highly relevant in understanding reaction dynamics catalysed by Au nanoparticles. 
The putative global minimum of Au$_{147}$ found by the authors using an ANN force field is reported in Fig. \ref{fig:jindal}, alongside the perfect icosahedron structure.
\cite{Jindal2017}
\\
The use of Gaussian process regression with 2-body, 3-body, and many-body descriptors (Eqs. \ref{eq:2bdescr}, \ref{eq:3bdescr}) and kernel functions (Eqs. \ref{eq:2bker}, \ref{eq:3bker}) was instead reported for the first time by Zeni et al., which modelled interatomic forces in Ni$_{19}$ nanoclusters. 
Thermodynamical properties as the melting point were probed thoroughly as the cost of the simulations carried out was 100000 times lower than DFT calculations. 
Fluxionality at temperature below melting was observed along such timescales.\cite{Zeni2018}
\\
\paragraph{Nanoalloys}
Also for the case of nanoalloys, machine learning force fields represent a highly helpful technology enabling thorough assessment of energetic and thermodynamic properties. 
\\
The first application of neural networks to build force fields of nanoalloys was carried on by Arthrith and Behler. 
They employed shallow artificial neural networks with Behler-Parrinello symmetry function descriptors (Eq. \ref{eq:ASF}) for the prediction of the composition and atomic ordering equilibrium architecture of AuCu alloy nanoparticles.
Site-based Monte Carlo simulations were used to sample the composition space while molecular dynamics simulations simultaneously enabled to sample the structure space. 
An extensive set of equilibrium properties for many temperatures and chemical potentials were thus assessed. 
Consistent with previous studies, the most stable structures were found to exhibit Cu(core)–Au(shell) configurations. 
Temperature dependent favourable alloy arrangement was also observed, with enhanced Au concentration at the particle core for increasing temperatures.
\cite{Artrith2015}
\\
Other alloys of Au where investigated by the groups of Bulusu, by means of neural network force fieldswith ASF as descriptors to predict global minimum structures of (AgAu)$_{55}$ nanoalloys across different compositions.
Pure Au and Au rich compositions minima resulted lower in energy compared to previous reports.
Thermodynamical and energetic properties were also thoroughly assessed 
(c-T phase diagram, surface area, surface charge, probability of isomers, and Landau free energies) to rationalize the enhancement of the catalytic property of Ag–Au nanoalloys by incorporation of Ag up to 24 by composition in AgAu nanoparticles. 
This result was found to match previous experimental data.
\cite{Chiriki2017b}
The development of novel methods for fast and accurate force and energy calculations is of even greater importance for the study of nanoparticles in conditions closer to the \emph{operando} one.
\\
High-dimensional neural network force fields using the ASFs were also incorporated with Monte Carlo and MD simulations by Kang et al. to identify not only active, but also electrochemically stable PtNiCu nanocatalysts for oxigen reduction reaction in acidic solution. 
The computationally efficient and precise approach proposed a promising oxigen reduction reaction candidate: a 2.6 nm diameter icosahedron comprising a 60 percent of Pt and a remaining equal mixture of Ni and Cu. \cite{Kang2018}
\\
\section{Conclusions}
\label{sec:conclustions}
In this review, we showcase how machine learning force fields offer the possibility to predict energies and forces with accuracy close to 
\emph{ab-initio} methods, but at a much lower computational cost.
The approaches to building machine learning force fields are multiple, and this review presents the major algorithms used so far in literature: linear regression, artificial neural networks and Gaussian process regression.
Their advent and exploitation allows to tackle the complexity of the energy landscape of metallic nanoparticles, even in the case of \emph{operando} conditions, for example when an oxide substrate is included and explicitely modelled.
Accurate but expensive algorithms to sample the free and potential energy landscapes of metallic nanoparticles (e.g. molecular dynamics, metadynamics, transition path sampling, basin hopping, harmonic approaches, nested sampling) will all greatly benefit from the deployment of machine learning-derived fast and accurate force fields.
Particular care must be placed on training database selection to ensure that the relevant parts of phase space are included without redundancy.
For this reason, a section of this review focuses entirely on describing the major algorithms for database selection.
\\
The core of this review is, however, to provide the state-of-the-art of machine learning force fields to model metallic nanoparticles and nanoalloys. 
Inspired by the on-going process of designing optimal metallic nanoparticles for target applications, and moving towards a numerical driven search, accurate estimates of how structural changes affect metallic nanoparticles' properties are in high need.
To achieve this result, fast and accurate force fields, that allow the exploration of long time scales without the caveat of fitted parameters, are required.
\\

\section*{Acknowledgements}
CZ and AG acknowledge funding by the Engineering and Physical Sciences
Research Council (EPSRC) through the Centre for Doctoral Training Cross Disciplinary Approaches to Non-Equilibrium Systems (CANES, Grant No. EP/L015854/1) and by the Office of Naval Research Global (ONRG Award No. N62909-15-1-N079). 
We are grateful to the UK Materials and Molecular Modelling Hub for computational resources, which is partially funded by EPSRC (EP/P020194/1).
FB acknowledges financial support from the UK Engineering
and Physical Sciences Research Council (EPSRC),
under Grants No. EP/GO03146/1 and No. EP/J010812/1.
KR acknowledges financial support from EPSRC, Grant No.
ER/M506357/1. 
KR and FB acknowledge the financial support offered by the Royal Society
under the project number RG120207.
\\

The authors would like to dedicate this review to their mentor, colleague, and friend prof. Alessandro ``Sandro" De Vita, passed away too early.

\bibliography{ml4mnp8_nodoi,FULL_2}

\end{document}